\documentclass[aps,amsfonts,amssymb,amsmath,groupedaddress,twocolumn]{revtex4}
\usepackage[english]{babel}

\begin{document}

\title{Using of small-scale quantum computers in cryptography with many-qubit entangled
states.}

\author{K.V. Bayandin}
\author{G.B. Lesovik}
\affiliation{L.D. Landau Institute for Theoretical Physics,
Russian Academy of Sciences, Kosygina Str. 2, 117940, Moscow}
\date{\today}

\begin{abstract}
We propose a new cryptographic protocol. It is suggested to encode
information in ordinary binary form into many-qubit entangled
states with the help of a quantum computer. A state of  qubits
(realized, e.g., with photons) is transmitted through a quantum
channel to the addressee, who applies a quantum computer tuned to
realize the inverse unitary transformation decoding of the
message. Different ways of eavesdropping are considered, and an
estimate of the time needed for determining the secret unitary
transformation is given. It is shown that using even small quantum
computers can serve as a basis for very efficient cryptographic
protocols. For a suggested cryptographic protocol, the time scale
on which communication can be considered secure is exponential in
the number of qubits in the entangled states and in the number of
gates used to construct the quantum network.
\end{abstract}

\maketitle

\section{Introduction}

In 1982 Feynman suggested that simulation of a quantum system
using another such system could be more effective than using
classical computers, which demand exponential time depending on
the size of the system~\cite{Feinman}. Later discussions focused
on the possibility of using quantum-mechanical systems for
solution of classical problems. For example, Deutsch's
algorithm~\cite{Deutsch} of verification of a balanced function
was the first quantum algorithm that worked more efficiently than
the classical analog.

The most famous of these,  Shor's  quantum factorizing
algorithm~\cite{Shor}, is capable of destroying widespread
cryptographic system RSA~\cite{RSA}. That fact made a strong
impression and speeded up the development of quantum
cryptography~\cite{book} and quantum information processing in
general.

It is important to note that quantum mechanics destroying
classical ways of coding still gives the possibility of
constructing new ones. At present, there exist many ways of coding
that use the quantum mechanics.

As an example, the quantum algorithm of key distribution using
orthogonal states should be mentioned~\cite{Bennet}. It was first
experimentally realized by Bennet and Brassard~\cite{Bennet_Exp},
who were able to carry out the transmission only at a distance of
forty centimeters. Later, a communication line of several
kilometers was realized~\cite{Geneva}.

Another example was first experimentally demonstrated in
1992~\cite{Entang}. The method uses pairs of entangled photons,
part of which, with the help of Bell inequalities of a special
form~\cite{Bell}, can be used to reveal attempted eavesdropping.

In the present article, another method of coding is proposed. It
uses quantum computers for creating entangled states of several
qubits. The safety of that method is based on the complexity of
tomography for those states.

Later, it will be convenient to treat a single qubit as a
spin-$\frac{1}{2}$ particle. To transmit information, Alice (the
sender) first transfers it into a set of units and zeros and
divides the numerals into groups of $K$ bits. Then, for every
group, she creates a set of $K$ spins in pure states. The spin
corresponding to a numeral gets the projection along the fixed
$Z$-axis if the numeral is zero and projected opposite to the axis
otherwise. After that, Alice employs a preset unitary
transformation $\hat{U}$ for every group of $K$ spins, thus
obtaining a set of entangled quantum-mechanical states that
hereafter will be called messages:
\begin{equation}
|\Psi_{k}\rangle=\hat{U}|k\rangle,
\end{equation}
where $|k\rangle$ is an unentangled state of spins with certain
projections along the $Z$-axis, and where the projections are
defined by the sequence of units and zeros for the binary record
of the number $k$.

Having received $K$ entangled spins, Bob (the receiver) employs
the inverse unitary transformation $\hat{U}^{-1}$, thus obtaining
the original separable state of spins with defined projections,
which can be measured and ,thereby, the secret message can be
decoded.

It is natural that only Alice and Bob know the unitary
transformation $\hat{U}$, providing that Eve (eavesdropper),
trying to measure the entangled quantum states, will obtain
probabilistic results defined by the quantum mechanics.

Further, we will consider the ways of learning how to decode the
transmitted information and, very importantly, how much time it
takes. We will consider two different ways: quantum tomography of
every entangled state and a simple guess of the quantum gate
network. The obtained results allow an estimate to be made of  how
long Alice and Bob may safely use the unitary transformation
without changing it.

\section{Quantum tomography of an entangled state.\label{s1}}

In the simplest case, Eve can determine the secret unitary
transformation if she knows exactly what information is sent by
Alice. We will not consider the question of how she can do that;
we will just assume that, having intercepted the message, Eve
knows exactly what information is encoded by Alice. Thus, for
simplicity, in this section we deal with many identical entangled
states.

The strategy for Eve is to employ quantum tomography for many
identical intercepted entangled states. In~\cite{Tomography} it
was shown that the density matrix of state of certain spins can be
derived without using quantum computers. The idea of the method is
based on a measurement of the probability
$p(\vec{n}_{1},m_{1};...;\vec{n}_{K},m_{K})$ for every spin
$\hat{s}_i$ projected into the state $m_{i}$ along the direction
$\vec{n}_{i}$. The density matrix is determined by the Monte-Carlo
integration
\begin{eqnarray}
&&\hat{\rho}_{calc}=\sum_{\{m_{i}\}=-\frac{1}{2}}^{\frac{1}{2}}\int...\int
\frac{d\vec{n}_{1}...d\vec{n}_{K}}{(4\pi)^K}p(\vec{n}_{1},m_{1};...;\vec{n}_{K},m_{K})\nonumber\\
&&\hat{K}_{S_1}(\vec{n}_{1},m_{1})...\hat{K}_{S_K}(\vec{n}_{K},m_{K}),\label{MonteKarlo}
\end{eqnarray}
where the kernel $\hat{K}_{S_i}(\vec{n}_{i},m_{i})$ acts in the
space of $i$-th spin.

Let us introduce distance in space of density matrices
\begin{equation}
|\hat{\rho}_{1},\hat{\rho}_{2}|=
\sqrt{\sum\limits_{i,j}|\hat{\rho}_{1}-\hat{\rho}_{2}|_{ij}^{2}},\;\;\;\;\;\;\;\;
|\hat{\rho}|= \sqrt{\sum\limits_{i,j}|\hat{\rho}|_{ij}^{2}}.
\end{equation}

It is known that, in the Monte-Carlo method, the relative
precision of integration converges as the inverse square of the
number of points used~\cite{MK}. In our case we have
\begin{equation}
\alpha=\frac{|\hat{\rho}_{calc}-\hat{\rho}_{true}|}{|\hat{\rho}_{true}|}\approx
\frac{1}{\sqrt{N}},
\end{equation}
where $N$ is the number of different sets of directions used for
measurement of spins.

Now we note that, for every set of fixed directions and for every
spin, it is necessary to measure all probabilities for every
combination of indices $\{m_{i}\}$. This takes about ${Const*2^K}$
intercepted messages.

Thus, we obtain that, in order to derive each density matrix with
precision $\alpha$ it is necessary to intercept
\begin{equation}
N_{intercepted}\approx Const*\alpha^{-2}*2^K\label{mainest}
\end{equation}
messages.

To compose the desired unitary transformation, Eve have to derive
the density matrices $\{\rho_{k}\}$ for all $2^K$ entangled
states. Every density matrix $\{\rho_{k}\}$ has a single
eigenvalue, $1$, and an eigenvector $|\Psi_{k}\rangle$
\begin{equation}
\hat{\rho}_{k}=|\Psi_{k}\rangle\langle\Psi_{k}|.
\end{equation}
Eve should find eigenvectors of the $2^K$ density matrices for all
entangled states and put them together; thus, she will get the
matrix $2^K \times 2^K$ for the unitary transformation $\hat{U}$
in the basis composed of vectors $|k\rangle$. Since the problem of
finding eigen vector for a matrix takes about $2^{2K}$ elementary
operations, the whole problem takes about
\begin{equation}
N_{operations}=2^{3K}
\end{equation}
operations, provided that we have a classical computer that can
operate with
\begin{equation}
N_{data}=2^{2K}
\end{equation}
complex numbers.

On top of this, for practical applications, Eve must construct a
quantum network by the unitary transformation. As we will see in
the next section, the number of necessary basic gates is
\begin{equation}
N_{gates}\approx 2^{2K}.
\end{equation}

Therefore, as Alice and Bob increase the number of bits contained
in a single message, the number of necessary intercepted messages,
the time necessary for deriving the unitary transformation and the
complexity of the constructed quantum network grow exponentially.

\section{Guessing of the unitary transformation\label{s2}}

Complicated unitary transformations can be constructed using
simple ones which mix states of one or two qubits. Examples of
actively studied gates for quantum networks are based on
superconducting circuits~\cite{Circuits}, resonant
cavities~\cite{Resonator}, linear ion traps~\cite{Ion} and nuclear
magnetic resonance~\cite{NMR}.

The operation of a quantum computer can be presented as a network
of sequential simple unitary transformations. The whole unitary
transformation has the form
\begin{equation}
\hat{U}=\hat{U}_{M}\hat{U}_{M-1}...\hat{U}_{2}\hat{U}_{1}.\label{q_net}
\end{equation}

Ekert and Jozsa showed~\cite{Ekert} that any unitary
transformation of qubits can be represented as a network of every
possible single-qubit gate and one type of double-qubit gate. An
example of double-qubit gate may can be "controlled NOT"\;, which
acts like $|a,b\rangle\rightarrow|a,a\oplus b\rangle$.

Due to the fact that every gate has its counterpart, which carries
out the inverse transformation, we can simply construct the
inverse transformation:
\begin{equation}
\hat{U}^{-1}=\hat{U}_{1}^{-1}\hat{U}_{2}^{-1}...\hat{U}_{M-1}^{-1}\hat{U}_{M}^{-1}.\label{q_net1}.
\end{equation}

Although the method of constructing of the quantum network by the
matrix of the unitary transformation was presented
in~\cite{Ekert}, in general case the algorithm requires a
polynomial number of gates over the dimension of the matrix
$\hat{U}^{-1}$, thus, in our case, it takes a number of gates that
is exponential in the number of qubits. Nevertheless, Alice and
Bob do not need to construct a quantum network to get a certain
unitary transformation: instead, they can just agree on a
particular one.

We assume that Alice and Bob possess identical quantum computers
which can carry out any of $L$ different simple unitary
transformations, provided that there exists an inverse
transformation for every one in the set. If Alice and Bob use the
simple transformations $M$ times, then the number of possible
quantum networks is
\begin{equation}
N_{quant}(L,M)=L^M\label{perebor}.
\end{equation}

Eve has no chance to guess the correct unitary transformation
trying every quantum network, taking into account that $M$ and L
should be greater than the square number of qubits $K^2$, because
Alice and Bob, a least, need to mix every qubit with each over.

As one can see, dependance (\ref{perebor}) is again exponential.
This formula yet does not take into account the fact that, for
every trial network, Eve must do several measurements of quantum
states to realize whether the network she has guessed is correct
or not. Let
\begin{equation}
p=|\langle k|\hat{U}_{guess}^{-1}\hat{U}|k\rangle|^2
\end{equation}
be the probability of erroneous acceptance of a trial unitary
transformation $\hat{U}_{guess}$ instead of the right one
$\hat{U}$. Then, the probability of not distinguishing this two
transformations after $n$ measurements is
\begin{equation}
P=p^n=e^{n\ln{p}}.
\end{equation}
Since, for overwhelming majority of quantum networks, the
probability $p$ is far less than one, a few measurements are
sufficient to realize that the network is erroneous.

As a result, we conclude that, to increase the security of the
cryptographic method, Alice and Bob should increase not only the
number of qubits but also the number of quantum gates used.

\section{The case of a priori known  time correlarions\label{corr}}

Earlier, we supposed that Eve knew what information was coded into
the entangled states. Now we will assume that she knows only time
correlations between messages of $K$ classical bits. The
correlations can be described by the value
\begin{equation}
\xi_{kl}(y)=\langle p_{k}(x)
p_{l}(x+y)\rangle_{x},\label{correlator}
\end{equation}
where $p_{k}(x)$ equals to unity, if $x$-th message is
$|k\rangle$, and zero otherwise.

We suppose that Eve possesses {\it a priory} information such as
the frequencies of appearance and the correlations between $K$-bit
messages that were sent by Alice. She tries then to construct a
quantum network that gives the same frequencies and correlations.

The estimated value of intercepted messages necessary for
deduction of the unitary transformation is divided into two parts:
the number of trial unitary transformations and the number of
necessary measurements for each of them to understand whether the
correlations are proper or not. The first part of the problem is
due to the entanglement, and the second is the same to the case of
classical replacement cipher.

The number of trial unitary transformations is defined by
formula~(\ref{perebor}). For calculation of the correlations, it
is necessary to measure a number of quantum states that is
polynomial in the value $2^K$
\begin{equation}
N_{cl}\approx P_{n}\left(2^K\right),\label{Nkl}
\end{equation}
where the power $n$ of the polynomial $P_{n}(x)$ corresponds to
taking into account of long time correlations.

The final number of messages to be intercepted is
\begin{equation}
N_{net}\approx N_{quant}*N_{cl}.\label{Nnet}
\end{equation}

\section{Discussion}

In the suggested way of encoding information, the number of
messages  that Eve must intercept is exponential in the number of
qubits and quantum gates used. This is clearly seen from
equations~(\ref{mainest}), (\ref{perebor}) and~(\ref{Nnet}).

According to the obtained estimations, it is necessary for Eve to
derive the structure of all $2^{K}$ entangled states, that is, to
intercept
\begin{equation}
N\approx C \times 2^{2K}
\end{equation}
messages. This corresponds to transmission of
\begin{equation}
N_{bit} \sim K\times 2^{2K}\label{v1}
\end{equation}
bits of classical information.

On the contrary, according to~(\ref{perebor}) it is necessary for
Alice and Bob to preset $M$ numbers less than $L$ to define the
order of simple unitary transformations. As we pointed earlier,
$M$ and $K$ are of order $K^2$, therefore, the number of bits
required for this is
\begin{equation}
N_{key}\sim K^2*\log_2{K^2}\label{v2}.
\end{equation}
This expression gives the length of the secret key that must be
shared by Alice and Bob. They can use a protocol of quantum key
distribution to get it. Expression~(\ref{v1}) shows how many
classical bits can be safely transmitted using that secret key.

Let us estimate the length of time that Alice and Bob may use a
given unitary transformation without changing it. For this, let us
consider the enciphering of telephone calls, which require
transmission of about fifty thousand bits per second. If the
quantum computer operates with $K=8$ qubits, then according to our
estimates, Eve should intercept $N\approx 65*10^3$ messages, so
Alice can send about $N*K=5*10^5$ classical bits or can talk to
Bob for ten seconds. If the computer operates with $K=16$ qubits,
then the time of guessing of the unitary transformation equals to
several weeks. And in the case of $K=24$ qubits the time of secure
conversation for Alice and Bob rises to four thousand years.

Although the suggested protocol requires a preset secret key, it
still has an advantage over classical block cipher algorithms,
which are also believed to be secure for transmission of an
exponential number of bits in the length of the key. The example
of RSA system and Shor's algorithm shows that quantum mechanics
can greatly simplify the breaking of codes based on complexity of
classical algorithms. On the contrary, the safety of the suggested
protocol is assured by fundamental laws of nature.

The main advantage of the suggested protocol is that Alice and
Bob, having arranged the secret transformation once, can use it
for a long time. The transmission is carried out in one direction,
as opposed to the protocols of secret key distribution, which
require repeated back-and-forth transmissions from Alice to Bob.

It should be mentioned that, according to the section \ref{corr},
the problem of determination of the secret unitary transformation
is added to the classical cryptographic problems. The main source
of additional security is the fact that the cloning of a state is
forbidden in any quantum-mechanical system~\cite{noncloning}. Due
to this theorem, a measurement in a wrong basis may give less
information than in the classical case, where an intercepted
message can readily be used for correlations calculation. In the
quantum case, a part of intercepted entangled states will be an
inevitable distraction for determination of the secret unitary
transformation.

Another issue is that, according to the noncloning
theorem~\cite{noncloning}, Eve destroys the quantum state
measuring it in a wrong basis, and, therefore, she is unable to
send the same state to Bob. In accordance with basic principles of
quantum cryptography~\cite{Bennet}, Bob can easily notice the
attempts of eavesdropping, and he can ask Alice to stop the
transmission. In another similarity to the case of relativistic
quantum cryptography~\cite{Molotkov}, Bob can detect the attempted
eavesdropping by the time delay for incoming messages.

Although the considered protocol looks promising, there are some
problems in its realization. First, it appears that the
construction  of quantum computers handling with tens of qubits is
still a matter of the future. Second, due to small decoherence
times for the systems with massive entangled particles, photons
remain the best objects for transmission of quantum states, but
the conversion of a state of qubits into a state of photons is a
challenging problem for experimentalists. Nevertheless, some
efforts have been made to study coupling between photons and
qubits~\cite{Schoelkopf} and to convert pairs of spin-entangled
electrons to pairs of polarization-entangled
photons\cite{Losslast}. Finally, during the transmission of
photons, there is the inevitable influence of the medium on their
states, and, therefore, the use of some quantum error-correction
techniques will be needed ~\cite{Knill}.

To conclude, we presented estimates showing that for the suggested
cryptographic protocol, the time that a secure secret unitary
transformation can be used is exponential in the number of qubits
within the entangled states and in the number of gates used to
construct the quantum network.

Although we can not at the moment present a rigorous proof of the
proper statements for Eve's general attack, the suggested protocol
in our opinion can serve as an interesting alternative to the
existing schemes in quantum cryptography. The main advantage of
the cryptographic protocol is that using even relatively small
quantum computers with several dozen qubits allows for a practical
scheme that is more efficient than existing ones in several
respects (e.g. weaker loading of communication channel).

We acknowledge discussions with S.\ Molotkov, G.\ Blatter, R.\
Renner, M.\ Feigelman, M.\ Skvortzov,  and financial support
through the Russian Science-Support Foundation, the Russian
Ministry of Science, and the Russian Program for Support of
Scientific Schools.

\end{document}